\documentclass[%
 reprint,
nofootinbib,
nobibnotes,
 amsmath,amssymb,
 aps,
prl,
]{revtex4-2}
\usepackage{svg}
\usepackage{graphicx}
\usepackage{dcolumn}
\usepackage{bm}
\usepackage{multirow}
\usepackage[english]{babel}
\usepackage{ulem}
\usepackage{natbib}
\usepackage[colorlinks, allcolors=blue]{hyperref}

\begin{document}
\newcommand{\x}{\bm{x}}
\newcommand{\y}{\bm{y}}
\newcommand{\er}{\bm{r}}
\renewcommand{\r}{\bm{r}}
\newcommand{\vzeta}{\bm{\zeta}}
\newcommand{\vi}{\bm{v}}
\newcommand{\D}{\mathcal{D}}
\newcommand{\msd}{\langle\r^2(t)\rangle}
\newcommand{\bphi}{\bar\phi}
\newcommand{\balp}{\bar\alpha}
\newcommand{\bs}{\bar s}
\newcommand{\half}[1]{\frac{#1}{2}}
\newcommand{\regnabla}[1]{\prescript{#1}{}{\nabla}}
\newcommand{\lt}{\left}
\newcommand{\rt}{\right}
\newcommand{\ttbl}{t_{tumbl}}
\newcommand{\cs}{g}
\newcommand{\fol}{f_o}
\newcommand{\ra}{r_a}
\newcommand{\dra}{\dot{r}_a}
\newcommand{\ddra}{\ddot{r}_a}
\renewcommand{\L}{\mathcal{L}}
\newcommand\T{\rule{0pt}{2.6ex}} 
\newcommand{\barf}{\bar f}
\newcommand{\bars}{\bar s}
\newcommand\B{\rule[-1.2ex]{0pt}{0pt}}
\newcommand{\tilalpha}{\tilde \alpha}
\newcommand{\at}[2]{\left. #1\right|_{#2}}
\newcommand{\ag}[1]{{\color{red}{\small\bf [AG: #1]}}}
\newcommand{\DE}{D_E}

\newcommand{\jr}[1]{{\color{blue}{\small\bf [JR: #1]}}}

\preprint{APS/123-QED}

\title{Anomalous diffusion and run-and-tumble motion of a chemotactic particle in low dimensions}

\author{Jacopo Romano}
\affiliation{SISSA --- International School for Advanced Studies and INFN, via Bonomea 265, 34136 Trieste, Italy}

\author{Andrea Gambassi}%
\affiliation{SISSA --- International School for Advanced Studies and INFN, via Bonomea 265, 34136 Trieste, Italy}

\date{\today}

\begin{abstract}
    We study the stochastic dynamics of a symmetric self-chemotactic particle and determine the long-time behavior of its mean squared displacement (MSD). The attractive or repulsive interaction of the particle with the chemical field that it generates induces a non-linear, non-Markovian effective dynamics, which results into anomalous diffusion for spatial dimensions $d \leq 2$. In one spatial dimension, we map the case of repulsive chemotaxis onto a run-and-tumble-like dynamics, leading to an MSD which, as a function of the elapsed time $t$, grows superdiffusively with exponent $4/3$.
    In the presence of attractive chemotaxis, instead, 
    the particle exhibits a slowdown, with the MSD growing logarithmically with time. In $d=2$, we find logarithmic aging of the diffusion coefficient, while in $d=3$ the motion reverts standard diffusive behavior with a renormalized diffusion coefficient.
\end{abstract}

\maketitle



\paragraph*{Introduction.}
Chemotaxis --- the ability to move in response to chemical gradients--- is a fundamental mechanism of motion 
in  biological and synthetic systems, governing processes such as bacterial and cellular migration~\cite{wadhams_making_2004,insall_steering_2022}, the motion of catalytic colloids and swimming droplets~\cite{michelin_self-propulsion_2023, jambon-puillet_phase-separated_2024,miura_ph-induced_2014,banno_mode_2013,jin_chemotaxis_2017, izri_self-propulsion_2014,blois_swimming_2021,romano2025dynamics}, and even the activity of individual enzymes~\cite{sengupta_enzyme_2013, yu_molecular_2009, dey_chemotactic_2014}. 
In these systems, the environment is continuously reshaped by the particles themselves, as they emit the chemical species that drive their own motion and create an evolving chemical landscape.
This self-chemotactic behavior represents a paradigmatic example of active matter~\cite{ramaswamy_active_2017}, 
and it has been extensively studied for its ability to induce self-propulsion when the particle maintains 
a chemical gradient along its surface~\cite{golestanian_propulsion_2005, michelin_self-propulsion_2023}. Janus colloids, i.e., particles engineered with asymmetric surface activity, represent a standard route to generate 
such gradients~\cite{golestanian_propulsion_2005, poggi_janus_2017,ebbens_direct_2011}, but this built-in asymmetry is not essential. 
In fact, spontaneous symmetry breaking can occur when the interaction between the colloid and the released chemical is repulsive: in this case, the chemical accumulation around the particle destabilizes its stationary configuration, resulting in autonomous propulsion. 
In the steady state and in the absence of thermal fluctuations, the particle moves with a constant speed that can be predicted analytically~\cite{michelin_spontaneous_2013, chamolly_stochastic_2019, morozov_self-propulsion_2019,picella_confined_2022}. 
However, because of their microscopic size, chemotactic particles are affected significantly by stochastic forces. Accordingly, in order to understand their transport properties, it is essential to characterize their mean squared displacement (MSD), defined as the mean $\langle\er^2(t)\rangle$ of the particle position $\er(t)$, as a function of time $t$. 
If the chemical produced by the colloid decays in time,  the temporal correlations of the velocity of the particle vanish rapidly and the the particle behaves diffusively at large times with modified diffusion coefficient~\cite{grima_phase_2006, grima_strong-coupling_2005}.
However, most chemotactic particles emit chemical species with negligible or no decay on the timescale of their dynamics.
In this case, modifications of the chemical landscape by the particle motion induce persistent, non-linear memory effects on its dynamics. Persistent memory typically leads to anomalous diffusion, as demonstrated in various low-dimensional systems, including trail-interacting particles~\cite{kranz_effective_2016}, self-interacting random walks~\cite{schulz_feedback-controlled_2005, amit_asymptotic_1983, obukhov_renormalisation_1983}, and passive tracers in active baths~\cite{granek_anomalous_2022}. Despite this, existing studies on low-dimensional chemotactic dynamics mostly focus on determining the effective diffusion coefficients~\cite{sengupta_dynamics_2009, izzet_tunable_2020, daftari_self-avoidant_2022,demery_perturbative_2011, dean_diffusion_2011}, leaving the precise functional form of the MSD at long times largely unexplored.
In this work, we aim at filling this gap by analyzing a paradigmatic model of a self-chemotactic particle and systematically studying the influence of the spatial dimensionality $d$ and the sign of the (attractive or repulsive) particle-field interaction $\alpha$ on their long-time dynamics. 
Using a scaling analysis, we identify a critical spatial dimension $d_c=2$ separating distinct dynamical regimes. For $d>d_c$, the particle exhibits normal diffusion at long times with renormalized diffusion coefficient. For $d\le d_c$, instead, we find anomalous diffusion: superdiffusion for repulsive interactions and subdiffusion for attractive interactions. To further characterize these behaviors, we map the dynamics onto simpler effective models -- a generalized Langevin equation with memory for the attractive case and a biased run-and-tumble model for the repulsive one. 
These mappings allow us to determine analytically the asymptotic behavior of the MSD, yielding logarithmic growth for the attractive case and superdiffusive growth scaling as $t^{4/3}$ in $d=1$.
These predictions are summarized in Tab.~\ref{tab:asymptotics}.

\begin{table}[t]
\caption{\label{tab:asymptotics}%
Long-time behavior of the MSD $\langle\er^2(t)\rangle$ depending on the dimensionality $d$ and on the sign of the interaction $\alpha$ with the field. $D(\cdots)$ is an unknown function.}
\begin{ruledtabular}

\begin{tabular}{c|ccc}
\multirow{2}{*}{chemotaxis} & \multicolumn{3}{c}{MSD} \\
& \textrm{$d=1$} & \textrm{$d=2$} & \textrm{$d=3$} \\[1mm]
\colrule
\\[-7.pt]
repulsive ($\alpha>0$) & $t^{4/3}$ &  $D(\ln t)\,t$ & \multirow{2}{*}{$t$} \\[1mm]
\\[-9.pt]
attractive  ($\alpha<0$) & $\ln t$ & $t/(\ln t)^2$  & \\

\end{tabular}
\end{ruledtabular}

\end{table}

\begin{figure}
    \centering
    \includegraphics[width=\linewidth]{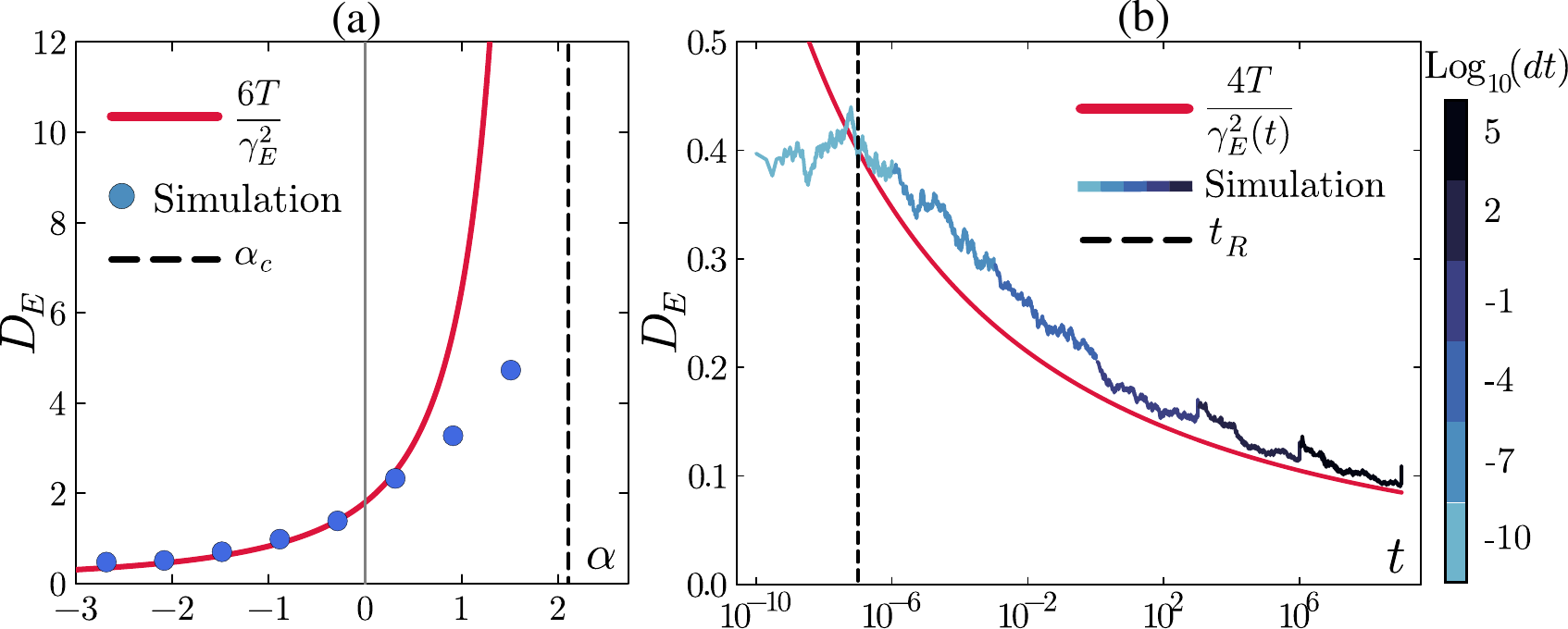}
    \caption{Effective diffusivity $\DE=\langle\r^2(t)\rangle/t$ as a function of (a) $\alpha$ in $d=3$ at large $t$ for $\D=1$, $t_R=10^{-3}$ and (b) $t$ in $d=2$ with $\alpha=-0.2$, $\D=0.5$, and $t_R=10^{-7}$. $\gamma=1$ in both cases.
    The red curves follow the Stokes-Einstein relation with effective viscosity $\gamma_E$ (Eq.~\eqref{eq:effDiff}), while blue symbols and lines are simulations of Eq.~\eqref{eq:langevin_selfchemo}. In panel (b) the time step is increased during the simulation to explore multiple scales \cite{supplementary}.
    %
    %
    }
    \label{fig:attractive_diffusivities}
\end{figure}

\begin{figure}
    \centering
    \includegraphics[width=.99\linewidth]{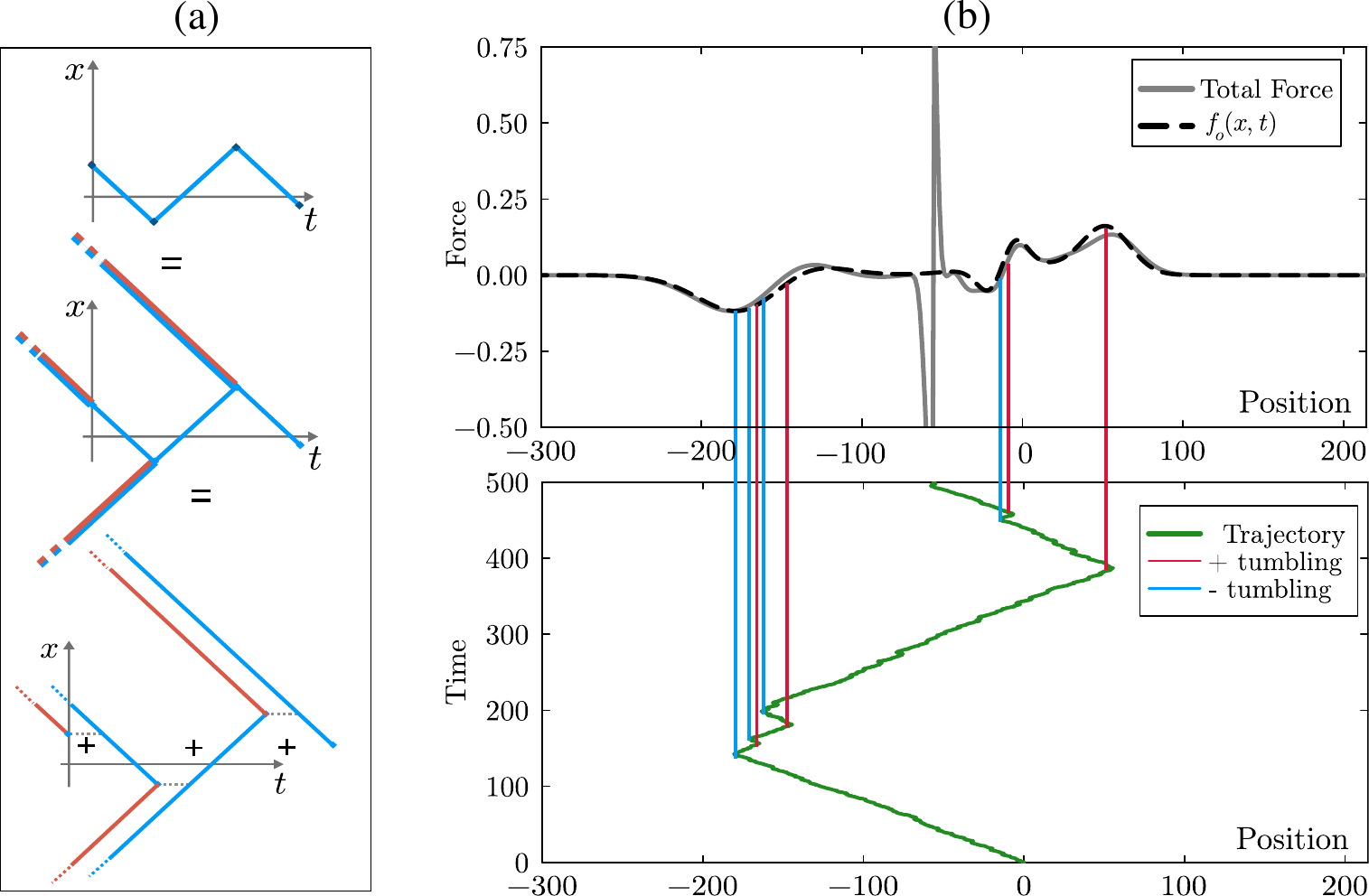}
    \caption{(a) The field generated by a run-and-tumble trajectory in $d=1$ (blue line, upper plot) can be decomposed as the sum of those generated by the segments of trajectories coming from $\pm\infty$ (central plot). Red trajectories generate a field with opposite sign than the blue and they can be paired as shown in the lower plot.
    (b) Top panel: force field $\alpha\nabla\phi(x,t)$ (solid line)
 from numerical integration of Eq.~\eqref{eq:reg_gradient} along a trajectory generated by simulating Eq.~\eqref{eq:langevin_selfchemo}, 
    compared with $f_{o}$ (dashed line) from Eq.~\eqref{eq:old_field}, as functions of $x$ at $t=500$. 
    Bottom panel: corresponding particle trajectory (green): it clearly emerges that tumblings are point sources for $\fol$.}
    \label{fig:traj_decomposition}
\end{figure}

\begin{figure}
     \centering
    \includegraphics[width=0.99\linewidth]{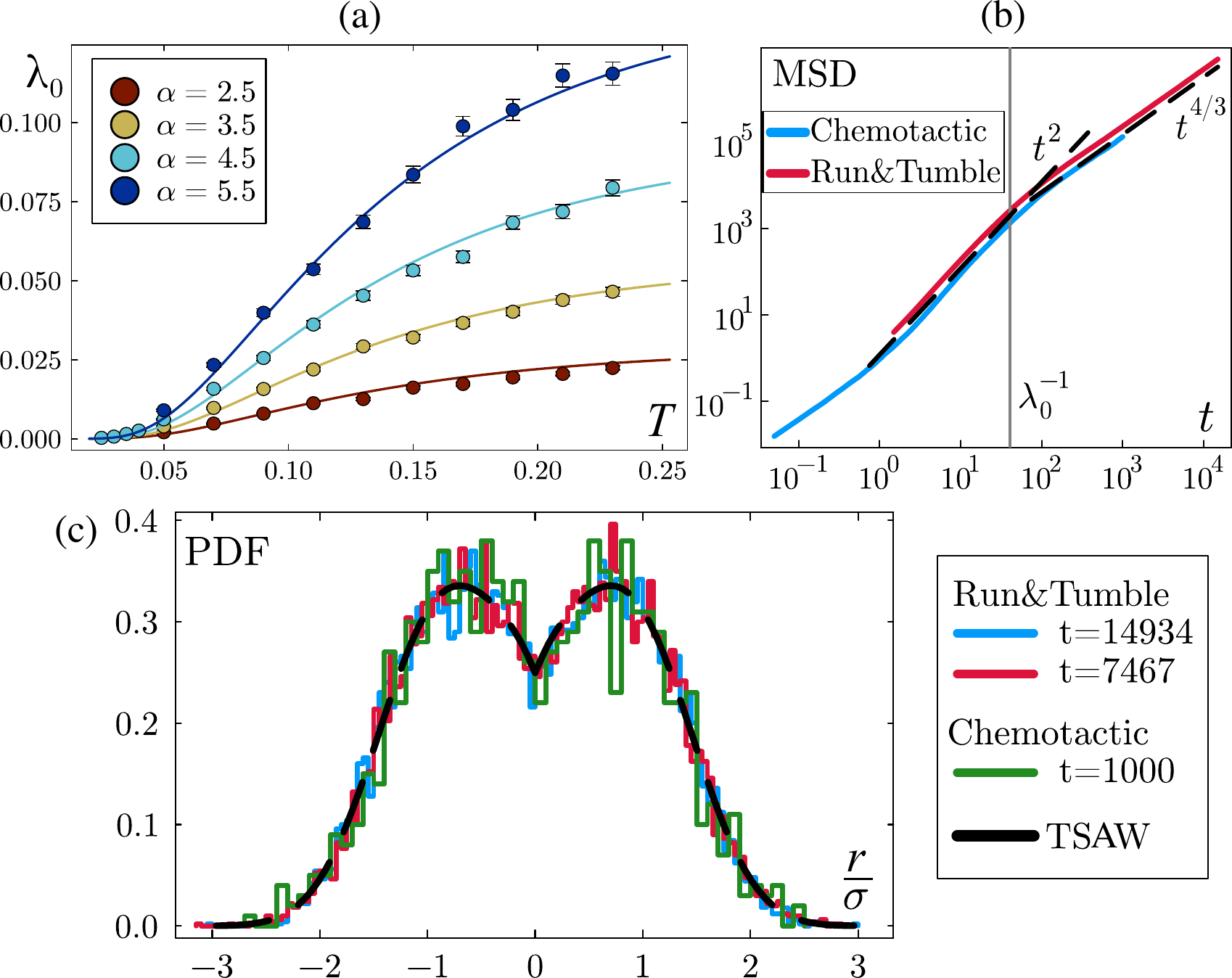}
     \caption{(a) $\lambda_0$ as a function of $T$ at varying $\alpha$. Symbols:  simulations of Eq.~\eqref{eq:langevin_selfchemo}, solid lines: fit to Eq.~\eqref{eq:tumblingrate}.
     (b) MSD  as a function of $t$ for the chemotactic (blue) and run-and-tumble dynamics (red). Both models display superdiffusive behavior $\propto t^{4/3}$ at long times and  ballistic behavior $\propto t^2$ at $t<\lambda_0^{-1}$ (vertical line). 
     (c) Empirical PDF of the rescaled position $r/\sigma(t)$, 
     for the chemotactic (green) and run-and-tumble (red and blue) dynamics, with $\alpha=3$, $T=0.12$, against analytical PDF for the  TSAW (dashed line) \cite{dumaz_marginal_2013}. See Ref.~\cite{supplementary} for details.
     }
     \label{fig:runandtumble}
 \end{figure}

\paragraph*{The model.} The dynamics of the system is determined by the evolution of the position $\er$ of the particle and of the concentration field $\phi(x,t)$ of the produced chemical. 
The particle, with overdamped dynamics, is subject to a phoretic force proportional to the gradient of the chemical field at its current position, to a viscous friction with coefficient $\gamma$, as well as to a stochastic force due to its interaction with a bath in equilibrium at temperature $T$. The field, instead, diffuses in space with a point source located at the colloid position, due to the chemical emission from the particle. 
The resulting equations are:
\begin{equation}
    \label{eq:langevin_selfchemo}
    \begin{cases}
    \gamma\dot \er(t)=-\alpha \nabla\phi_{t_R}(\er(t),t)+\vzeta(t),\\[1mm] 
    \partial_t\phi(\x,t)=\D\nabla^2\phi(\x,t)+\delta^d(\x-\er(t)),
    \end{cases}
\end{equation}
where $\D$ is the diffusivity of the chemical field, 
$\alpha$ 
the phoretic constant, and 
the noise $\vzeta(t)$ has zero mean and variance $\langle\zeta_i(t)\zeta_j(t')\rangle=2T\delta(t-t')\delta_{ij}$.
%
A similar set of equations appears in diffusion through random media, studied for both passive \cite{demery_non-gaussian_2023, demery_thermal_2011} and active \cite{demery_perturbative_2011,dean_diffusion_2011} particles. Here, however, chemical production and the absence of decay leads to anomalous diffusion at all parameter values.
Note that, in the presence of a point source, $\nabla\phi$ actually diverges at the particle position for $d>1$. Accordingly, one needs to regularize the divergence: this introduces a timescale $t_R$, indicated as a subscript of $\phi$ in Eq.~\eqref{eq:langevin_selfchemo}, which is assumed to be smaller than any other scale of the system. 
Different regularization schemes introduced in the literature include smoothing the source term in Eq.~\eqref{eq:langevin_selfchemo} with a Gaussian~\cite{demery_non-gaussian_2023,daftari_self-avoidant_2022} (accounting for the finite size of the colloid) or computing the field produced by the particle only up to time $t'<t$~\cite{sengupta_dynamics_2009} (accounting for the time delay between the production of chemical and the phoretic response). Changing regularization scheme does not alter the dynamics (up to numerical factors), provided that the scheme itself does not break the rotational symmetry of the problem. 
We regularize Eq.~\eqref{eq:langevin_selfchemo} as follows: being $G_d(\x,t)=(4\pi \D t)^{-d/2} e^{-x^2/(4\D t)}$ the solution of the diffusion equation, the field generated at position $\x$ by the colloid 
with trajectory $\{\er(t'), 0\le t'\le t\}$ is $\phi(\x,t)=\int_0^t {\rm d} t' G_d(\x-\er(t'),t-t')$. Then we define:
\begin{equation}
    \label{eq:reg_gradient}
\phi_{t_R}(\x,t)=\int_0^t {\rm d} t' G_d(\x-\er(t'),t+t_R-t'),
\end{equation}
i.e., the field $\phi$ at position $\x$ but at a later time $t+t_R$, assuming that no chemical is released from $t$ to $t+t_R$.
Considering $\phi_{t_R}(\x,t)$ for $t'\simeq t$, it can be shown that $\nabla\phi_{t_R}(\x,t)$ 
diverges at $\x=\r(t)$ as $t_R\rightarrow 0$ for $d\geq2$. In particular, this divergence is proportional to the particle velocity: $\nabla\phi_{t_R}(\r,t)\rightarrow\cs(t_R)\dot \r(t)+ \mbox{finite}$, where the finite part is independent of $t_R$, while $g$ is given by:
\begin{equation}
\label{eq:core_struct_func}
    g(t_R)=\begin{cases}
        (\ln t_R) / (8\pi \D^2)&\quad \text{for} \quad d=2,\\
        -1/(12\pi^{3/2} \D^2\sqrt{\D t_R}) & \quad \text{for} \quad d=3.
    \end{cases}
\end{equation}
\paragraph*{Scaling analysis.} To study the long-time behavior of the system we perform a scaling analysis of Eq.~\eqref{eq:langevin_selfchemo} by applying the transformations $\er\rightarrow b\er$ and $\x \rightarrow b\x$ to the space variables with a scale parameter $b\simeq1$. Time, field, and the phoretic constant are correspondingly scaled as $t\rightarrow b^z t$, $\phi\rightarrow b^{d_{\phi}}\phi$, $\alpha\rightarrow b^{d_{\alpha}}\alpha$, where $z$, $d_{\phi}$, and $d_{\alpha}$ are determind such that Eq.~\eqref{eq:langevin_selfchemo} is invariant under the scale transformation. 
Note that 
the scale transformation reduces the cutoff $t_R$ to $t_R/b^z$. We use Eq.~\eqref{eq:core_struct_func} to restore the original cutoff, by noticing that for $b\simeq 1$ one has
$\nabla\phi_{t_R/b^z}=\nabla\phi_{t_R}-zt_R\frac{\partial g}{\partial t_R}\dot \r(t)\ln b$. The rescaled equations of motion are:
\begin{equation}
    \label{eq:scaled_selfchemo}
    \begin{cases}
    \gamma b^{1-z}\dot \r=- b^{d_{\alpha}+d_{\phi}-1}[\alpha\nabla\phi_{t_R}-\beta_{\gamma}\dot \r\ln b]+b^{-z/2}\vzeta(t)\\
        b^{d_{\phi}-z} \partial_t\phi(\x,t)=b^{d_{\phi}-2}\D\nabla^2\phi(\x,t)+b^{-d}\delta^d(\x-\r(t)).
    \end{cases}
\end{equation}
with $\beta_{\gamma}(t_R)=\alpha z\frac{\partial g}{\partial \ln t_R}$ for $d=3$ or 2 and $\beta_{\gamma}=0$ for $d=1$.
Imposing that Eq.~\eqref{eq:scaled_selfchemo} is the same as Eq.~\eqref{eq:langevin_selfchemo} requires
$z=2$, $d_{\phi}=2-d$, $d_{\alpha}=-d_{\phi}$, 
but this is not sufficient. We also transform the viscosity as $\gamma\rightarrow\gamma+\beta_{\gamma}(t_R)\ln b$ to account for the short-scale correction to the gradient due to the change in cutoff. 
Since Eq.~\eqref{eq:langevin_selfchemo} is unchanged under the aforementioned transformations, also $\langle\r^2\rangle$ has to be unchanged. Imposing its invariance for $b\simeq1$ gives the Callan-Symanzik equation:
\begin{equation}
    \label{eq:CS}
    \left[-2+\alpha d_{\alpha}\partial_{\alpha}+zt\partial_t +\beta_{\gamma}(t_R)\partial_{\gamma}\right]\langle\r^2\rangle=0.
\end{equation}
Using the method of characteristics, we solve Eq.~\eqref{eq:CS} to express the MSD as $\msd=\DE(\alpha_E(t),\gamma_E(t),t_R)\,t$, where the 
effective diffusivity $D_{E}$ depends on the effective coupling $\alpha_E(t)=\alpha \left(t/t_R\right)^{-d_\alpha/z}$ and viscosity: 
\begin{equation}
    \label{eq:effDiff}
    \gamma_E(t)=\begin{cases}
        \gamma-\dfrac{\alpha}{8\pi\D^2}\ln \left(t/t_R\right)&\text{ for }d=2,\\
        \gamma-\dfrac{\alpha}{12\pi^{3/2} \D^2\sqrt{\D t_R}}&\text{ for }d=3.
    \end{cases}
\end{equation}
For $d<2$, $|\alpha_E|$ 
 grows with time  
 and thus the MSD depends on the asymptotic form of $\DE$ at large $\alpha$, which is unknown from scaling alone. This case will be discussed separately below. For $d\geq 2$ and 
 $\alpha<0$, Eq.~\eqref{eq:effDiff} shows how the chemical cloud surrounding the particle generally slows down its dynamics by increasing the effective viscosity $\gamma_E$(t) of the medium compared to its bare value $\gamma$.
Moreover, as for $d>2$ $\alpha_E$ vanishes at long times,
 $\DE$ is determined by the Stokes-Einstein relation  $\DE(0,\gamma_E,t_R)=2 d T/\gamma_E^2$ (for $d=2$  this is an approximation which requires small $\alpha$).
In particular, for $d=3$, only the chemical produced at times close to $t$ affect the dynamics of the colloid and thus the effective viscosity approaches a constant value which is function of the bare coupling $\alpha$. This is shown in Fig.~\ref{fig:attractive_diffusivities}(a). 
 By contrast, in $d=2$, the dynamics is affected by the chemical produced at early times, causing a logarithmic growth of the effective viscosity $\gamma_E$ and thus ageing of the system. The corresponding reduction in the effective diffusivity upon increasing time is shown in 
Fig.~\ref{fig:attractive_diffusivities}(b).
 Finally, at $\alpha>0$ the Stokes-Einstein relation does not correctly capture the dependence of the diffusivity on $\alpha$, as $\gamma_E$ vanishes at $\alpha=\alpha_c = 12\pi^{3/2} \D^2\sqrt{\D t_R}\gamma$ (in $d=3$) or $\alpha>0$ (in $d=2$), causing a divergence of $\DE$. 
This change of sign 
corresponds to the onset of the self-propelling regime of the chemotactic particle, similarly to other active systems~\cite{pei_transfer_2025,maes_fluctuating_2020, pei_induced_2025,granek_anomalous_2022}. 
%
%
This implies that for $\alpha > 0$ and $d = 2$, the precise dependence of $\DE$ on $\ln t$ remains undetermined. The numerical investigation of this case is also difficult, due to the long timescales required to resolve a logarithmic growth, and it is left for future works.
For $d>2$ and $\alpha>0$, instead, the scaling analysis still predicts a diffusive behavior at long times, although with an undetermined diffusion coefficient.
\paragraph*{Dynamics in one dimension.} In $d=1$ the chemical interaction generates non-Markovian effects which 
affect more strongly the 
particle dynamics. 
Note that, now, we 
can 
set $t_R = 0$ without encountering divergences. Without loss of generality, we also set $\mathcal{D}=1$ and $\gamma=1$ 
The dependence on $\mathcal{D}$ and $\gamma$ can be recovered by applying the transformations: $\alpha\rightarrow\alpha/(\gamma \mathcal{D}^2)$, $T\rightarrow T/(\gamma^2 \mathcal{D}^2)$, $t\rightarrow t/\mathcal{D}$, $\phi\rightarrow \phi/\mathcal{D}$
.
Since 
for $d<2$ 
the scaling analysis is inconclusive, a separate study of the attractive and repulsive interactions is required.
Consider, first, the attractive case $\alpha<0$: as a subdiffusive growth of the MSD is expected, the displacement $[r(t)-r(t')]^2$ grows slower than the elapsed time $t-t'$ if the latter is large and/or $T$ is low. 
Accordingly, we can approximate:  $\nabla G_1(x,t)\simeq -\frac{1}{4\sqrt{\pi}}\frac{x}{t^{3/2}}$ in the expression of $\nabla \phi_{t_R=0}$ following from Eq.~\eqref{eq:reg_gradient}, to linearize the memory kernel induced by the self chemotaxis. This leads to the generalized Langevin equation for the colloid:
\begin{equation}
\label{eq:genlangevin}
    \dot r(t)-\frac{\alpha}{2 \sqrt{\pi}}I_t[\dot r]= \zeta(t),
\end{equation}
where $I_t$ is an effective retarded friction due to chemotaxis, given by
$I_t[f]=\int_0^tdt'\left[(t-t')^{-1/2}\!-\!t^{-1/2}\right]f(t')$. 
The solution of Eq.~\eqref{eq:genlangevin} (see Ref.~\cite{supplementary} for details) implies, at long times,
\begin{equation}
    \label{eq:MSDlargetimesattractive}
    \langle r^2(t)\rangle \simeq\frac{8\pi T}{\alpha^2 a^2(0)}\ln t+c,
\end{equation}
where
$a(0) = 1.39\ldots$ 
%
while $c$ is an undetermined constant. Numerical simulations confirm this prediction \cite{supplementary}.
In the repulsive case $\alpha>0$ both nonlinearity and memory 
play a crucial role in the dynamics. 
In the noiseless limit $T \to 0$, the particle exhibits self-propulsion: in fact,  by inserting the ansatz $r(t) = \pm vt$ into Eq.~\eqref{eq:langevin_selfchemo}, we find two steady-state solutions with speed $v = \alpha/2$, introducing a natural timescale $\tau_\alpha = 4/v^2 = 16/\alpha^2$. 
At small but finite temperature $T$, the noise induces tumbling events between these two states, inverting the direction of motion. When a tumble occurs, the particle visits again previously explored regions and interacts with the residual chemical field it left behind. This self-generated chemical memory modifies the dynamics and plays a key role in determining its long-time behavior. To investigate this, one needs to understand both the mechanism of tumbling and how it is influenced by the chemical memory. 
Note that these tumblings --- defined as changes in the sign of the velocity --- cannot be straightforwardly identified from $\dot{r}(t)$, which is non-differentiable due to the white noise. 
To single out the slow dynamics, we project the trajectory onto low-frequency modes by convolving Eq.~\eqref{eq:langevin_selfchemo} with the exponential kernel $g(t) = \tau^{-1} e^{-t/\tau}$.  In particular, we define the coarse-grained position as $r_s(t) = (g * r)(t)$, where $(g * f)(t) = \int_{-\infty}^t g(t - t') f(t')$. We choose the convolution timescale to be $\tau \propto \tau_\alpha$. 
We assume that fast fluctuations average out and we approximate the convoluted chemotactic force as $g * \nabla \phi(r(t), t) \simeq \nabla \phi(r_s(t), t)$. Under this approximation, $r_s(t)$ evolves according to the same dynamics as $r(t)$ in Eq.~\eqref{eq:langevin_selfchemo}, but with the white noise $\zeta(t)$ replaced by the smoothened noise 
$\zeta_s(t) = g * \zeta(t)$.
This filtering allows one to identify the tumbling events from the slow variable $r_s(t)$.
To understand how the tublings occurr, we split the chemotactic force $-\alpha \nabla \phi(r_s(t), t)$ in Eq.~\eqref{eq:reg_gradient} into two contributions: one arising from the chemical produced since the last tumble and the other from those released before that time, which we denote by $\fol(r_s(t),t)$. 
Note that if $T$ is sufficiently low, tumblings are rare and the chemical field has time to relax in the meanwhile. 
This causes $\fol$ to be a slowly varying function, consistent with the requirement that $r_s$ is also slow. For infinitely slow $\fol$ and $\zeta_s$, 
the systems admits an adiabatic solution with velocity $\dra(t)=\epsilon \, \alpha/2 +\fol(r_s(t),t)+\zeta_s(t)$, where $\epsilon=\pm1$ indicates the direction of the self-propelling motion. Expanding $r_s(t)=\ra(t)+\delta r(t)$ and Laplace-transforming the equation for $\delta r$ we arrive to \cite{supplementary}:
\begin{equation}
    \label{eq:lap_transf}
    \L[\delta r](s)=\frac{\alpha\L\lt[\ddra/v_a^3\rt](s)}{\lt[s+\frac{\alpha}{2} v_a-\frac{\alpha}{4}\lt(\frac{2v_a^2+4 s}{\sqrt{v_a^2+4s}}\rt)\rt]},
\end{equation}
where $\L$ denotes the Laplace transform and $v_a=|\dra(t)|$. The tumbling instability happens when at least one pole of $\L[\delta r](s)$ 
has a positive real part, leading to exponential growth of $\delta r(t)$. According to Eq.~\eqref{eq:lap_transf} this requires that the particle is slowed down to $v_a<(\sqrt{5}-2)\alpha$ and therefore that 
$-\epsilon[ \fol(r_s,t)+\zeta_s(t)]$ exceeds the critical force $f_c=(\sqrt{5}-5/2)\alpha$.  
%
%
The rate $\lambda^{\epsilon}(\fol)$ with which fluctuations in $\zeta_s$ cause $\fol+\zeta_s$ to satisfy this tumbling condition are given by \cite{supplementary}:
\begin{equation}
    \label{eq:tumblingrate}
    \lambda^{\epsilon}(\fol)\simeq \frac{\alpha(f_c+ \epsilon \fol)B}{4\sqrt{2\pi T A}}\exp{\lt[- 8A\frac{(f_c +\epsilon \fol)^2}{\alpha^2T}\rt]}.
\end{equation}
Here, $A$ and $B$ are two fitting parameters: $A$ optimizes the scale separation between fast and slow dynamics (thus fixing the coarse-grained timescale $\tau = A \tau_\alpha$),
 while $B$ accounts for sub-leading corrections  of order $O(T)$ in $f_c$, $v_a$, and $\tau_\alpha$.
 %
 %
 Their values are determined by measuring the rate $\lambda_0=\lambda^{\pm}(0)$  of the first tumbling from simulations of Eq.~\eqref{eq:langevin_selfchemo} at various $\alpha$ and $T$ and fit Eq.~\eqref{eq:tumblingrate}.
 %
 %
 %
 %
%
To conclude the derivation of the effective run-and-tumble dynamics we express $\fol(x,t)$ in terms of the positions at which previous tumblings occurred. As at low $T$ the separation between two consecutive tumblings is large compared to the typical length scale $l=\D/v$ of the dynamics, we can study $\fol$ in the limit $l\rightarrow 0$.
We start by considering the gradient $\partial_x\phi_\pm$ of the chemical field $\phi_\pm$ produced by a colloid moving with constant velocity $\pm v$ up to time $t$. 
As $\phi_\pm$ changes significantly only close to the particle position 
$r(t)$, at large scales one has $\partial_x\phi_{\pm}\simeq\mp l^{-1}\delta(x-r(t))$, see Ref.~\cite{supplementary}.
The chemical field generated by a run-and-tumble trajectory, with tumblings occurring at the space-time points $(x_i, t_i)$, can then be decomposed as illustrated in Fig.~\ref{fig:traj_decomposition}(a). For each tumbling, one adds (blue) and subtracts (red) the field $\phi_{-}$ or $\phi_+$ due to a fictitius particle reaching $x_i$ from $|x|= \infty$ by moving with a velocity opposite to that of the trajectory right before the tumbling.
By pairing these contributions, one finds that the $i$-th tumbling generates a field $\epsilon_i(\phi_+ - \phi_-)$, where $\epsilon_i$ indicates the sign of the velocity along the trajectory immediately before the tumbling. This field acts as a point source of chemical gradient, causing a chemotactic force field $f_i(x,t) = 4\epsilon_i G_1(x - x_i, t - t_i)$. 
%
%
$\fol$ is then given by the sum of the contributions due to the various tumblings, plus the one of the unpaired trajectory from the initial condition, 
i.e.,
\begin{equation}
\label{eq:old_field}
\fol(x,t) = 2 \epsilon_1 G_1(x,t) + 4 \sum_{i=1} \epsilon_i G_1(x - x_i, t - t_i).
\end{equation}
%
%
Equations~\eqref{eq:tumblingrate} and \eqref{eq:old_field} define an effective run-and-tumble dynamics influenced by the chemotactic memory, which we simulate and compare with the original dynamics, as shown in Fig.~\ref{fig:runandtumble}. 
In particular, by fitting the measured $\lambda_0$ with Eq.~\eqref{eq:tumblingrate}, we obtain $A=0.42$ and $B=8.05$. 
 The quality of the resulting fit is shown in Fig.~\ref{fig:runandtumble}(a) at varying $T$ and $\alpha$.
The MSD  as a function of $t$ and the probability distribution (PDF) of $r/\sigma(t)$ ($\sigma^2(t)=\langle r^2(t)\rangle$) at two (large) $t$,
resulting from the effective run-and-tumble dynamics and from the chemotactic Eq.~\eqref{eq:langevin_selfchemo} are compared in Figs.~\ref{fig:runandtumble}(b) and \ref{fig:runandtumble}(c), respectively, showing good agreement. 
In particular, the MSD for both models show the same scaling exponent and PDF at long times, which we now derive for the latter. 
Let $P(r,t)$ denote the probability of finding the chemotactic particle at 
position $r$ and time $t$ and $\barf(r,t)=\langle \fol(r,t)\rangle$ 
%
the force acting on the particle, averaged over all possible past trajectories.
Within the mean-field approximation and for large $t$ they satisfy \cite{supplementary}:
\begin{equation}
\label{eq:FP_eff}
\begin{cases}
    \partial_t P(r,t)=\dfrac{\alpha^2}{8 \lambda_0}\partial^2_r P(r,t)+\dfrac{\alpha\lambda_1}
    {4\lambda_0}\partial_r[\barf(r,t)P(r,t)],\\[2mm]
\bar{f}(r,t) =-\alpha \int_0^t dt' \!\int \!dr' P(r',t') \partial_r G_1(r - r', t - t').
\end{cases}
\end{equation}
with $\lambda_1=\frac{\partial \lambda^+(f)}{\partial f}\big|_{f=0}$.
In the first equation, the diffusion term stems from the run-and-tumble motion and competes with the advective term due to $\fol$. 
Using the ansatz $P(r,t) = t^{-z} U(r/t^z)$ in Eq.~\eqref{eq:FP_eff} with $z>1/2$, one 
finds $z = 2/3$, such that $\langle r^2(t) \rangle \propto t^{4/3}$ \cite{supplementary}. 
This prediction is confirmed by the numerical simulations of both the chemotactic model in Eq.~\eqref{eq:langevin_selfchemo} and the effective run-and-tumble dynamics given by Eqs.~\eqref{eq:tumblingrate} and~\eqref{eq:old_field},  see Fig.~\ref{fig:runandtumble}(b).
Remarkably, this exponent coincides with that one of the ``true" self-avoiding random walk (TSAW)~\cite{amit_asymptotic_1983,pietronero_critical_1983,schulz_feedback-controlled_2005}, a random walk in which the probability of visiting a site decreases with the number of previous visits to it.
 In fact, since the motion of the particle is superdiffusive, at long times the spreading of the chemical by diffusion is negligible compared to the typical particle displacement: the chemical is effectively frozen at its emission points and this fact turns the run-and-tumble into a TSAW. This is confirmed by the agreement between the PDF obtained from both models and the analytical solution of the TSAW~\cite{dumaz_marginal_2013,toth_true_1998}, see 
Fig.~\ref{fig:runandtumble}(c). 
\paragraph*{Conclusions.} 
We have shown that non-Markovian, chemically mediated interactions cause anomalous diffusion of self-chemotactic particles for  $d \le 2$.
This observation has profound implications for the collective behavior of chemically active matter \cite{saha_clusters_2014, theurkauff_dynamic_2012}. While most theoretical descriptions of chemotactic particles rely on the assumption of fast chemical diffusion $\D\rightarrow\infty$ \cite{mahdisoltani_nonequilibrium_2021,zinati_stochastic_2021},
this assumption breaks down for $\alpha>0$ and $d\le 2$, as particles become superdiffusive. In this case, the relevant scaling limit is $\D\rightarrow0$, suggesting that collective chemotaxis in low dimensions may belong to a fundamentally different universality class than its higher-dimensional counterpart.
%
%
Understanding these emergent behaviors remains an open and compelling challenge.
An interesting generalization of our work involves replacing the production of chemical in Eq.~\eqref{eq:langevin_selfchemo} with an interaction which conserves its overall mass, encompassing the case of a probe non-reciprocally coupled to a bath \cite{benois_enhanced_2023}.
Finally, another direction is to consider statistical properties  beyond transport, for example examining how memory influences first-passage properties, which are central to search and sensing strategies in active systems.


\bibliographystyle{apsrev4-2} 
\bibliography{bibliography.bib}
\onecolumngrid
\clearpage

\begin{center}
\textbf{\large Supplementary material for:\\
Anomalous diffusion and run-and-tumble motion of a chemotactic particle in low dimensions}\\[1em]
Jacopo Romano and Andrea Gambassi\\
{\it SISSA --- International School for Advanced Studies and INFN, via Bonomea 265, 34136 Trieste, Italy}
\end{center}
\vspace{1cm}
\setcounter{equation}{0}
\setcounter{figure}{0}
\renewcommand{\theequation}{S\arabic{equation}}
\renewcommand{\thefigure}{S\arabic{figure}}
\renewcommand{\thesection}{S.\Roman{section}}

This supplemental material provides details concerning the derivation of multiple technical results stated in the main text, as well as on the algorithms used in the simulations. 

\section{Derivation of the MSD and simulations for the attractive case in $d=1$}
\label{app:1dattractive}

In the attractive case, the self-chemotactic dynamics~\eqref{eq:langevin_selfchemo} can be approximated by the generalized langevin equation~\eqref{eq:genlangevin}, allowing for analytical progress.  In order to derive the expression of the mean-squared displacement (MSD) in Eq.~\eqref{eq:MSDlargetimesattractive} we use the following representation of the delta-function:
\begin{equation}
    \label{eqsupp:deltarepr}
    \delta(t_1-t_2)=\int_{-\infty}^{+\infty}\frac{du}{2\pi}t_1^{iu-1/2}t_2^{-iu-1/2},
\end{equation}
with $t_1$, $t_2>2$. This expression is obtained by doing a change of variable, which allows one tho write   
%
%
$\delta(t_1-t_2)=t_1^{-1/2}t_2^{-1/2}\delta(\ln t_1 - \ln t_2)$. Then, representing $\delta(\ln t_1 -\ln t_2) $ via its Fourier transform, i.e.,  $\delta(x)=1/(2\pi)\int \!du\, e^{iux}$, one arrives at Eq.~\eqref{eqsupp:deltarepr}.

We now derive the two-time velocity correlation function $\langle \dot r(t_1)\dot r(t_2)\rangle$ by substituting the formal solution $\dot r(t)=\{1-\tilalpha I_t\}^{-1}[\zeta]$ of Eq.~\eqref{eq:genlangevin} in the correlator, where we introduced $\tilalpha=\alpha/(2\sqrt{\pi})$ and the inverse is understood in the sense of operators acting on functions of time. 
Using Eq.~\eqref{eqsupp:deltarepr}, the linearity of $I_t$, 
 and averaging over the white noise we find:
\begin{equation}
    \label{eqsupp:velcorr1}
        \langle\dot r(t_1)\dot r(t_2)\rangle=2 T\int_{-\infty}^{+\infty}\frac{du}{2\pi}\{1-\tilalpha I_{t_1}\}^{-1}[t_1^{iu-\frac{1}{2}}]\{1-\tilalpha I_{t_2}\}^{-1}[t_2^{-iu-\frac{1}{2}}].
\end{equation}
Although the action of $I_t$ [see its definition after Eq.~\eqref{eq:genlangevin}] on a generic function cannot be given an explicit expression, this can be done on monomials. Specifically, one finds $I_t[t^p]=a(p+1)t^{p+1/2}$, where
\begin{equation}
    \label{eqsupp:ap}
    a(p)=\sqrt{\pi}\frac{\Gamma(p)}{\Gamma(p+1/2)}-\frac{1}{p},
\end{equation}
and $\Gamma$ denotes the Gamma function. 
Conversely, the action of $I_t^{-1}$ on monomials is given by: $I_t^{-1}[t^p]=\/a(p+1/2)t^{p-1/2}$. Moreover, for large $t\gg\alpha^{-2}$ we have $\tilalpha I_t[t^p]\gg t^p$, and therefore we can approximate $[1-\tilalpha I_t]^{-1}\simeq \/\tilalpha I_t^{-1}$. 
Using the latter in Eq.~\eqref{eqsupp:velcorr1}, we find:
\begin{equation}
    \label{eqsupp:velcorr2}
    \langle\dot r(t_1)\dot r(t_2)\rangle=\frac{8\pi}{ \alpha^2}\int_{-\infty}^{+\infty}\frac{du}{2\pi}\frac{t_1^{iu-1}t_2^{-iu-1}}{|a(iu)|^2}.
\end{equation}
To evaluate the aysmptotic behaviour of the MSD we integrate both $t_1$ and $t_2$ in Eq.~\eqref{eqsupp:velcorr2} from $t_c$ to $t$, where $t_c$ is a cutoff time of order $\alpha^{-2}$, before which the large-$t$ approximation is no longer valid. This leads to:
\begin{equation}
\label{eqsupp:almostmsd}
    \langle r^2(t)\rangle=\frac{8\pi}{ \alpha^2}\int_{-\infty}^{+\infty}\frac{du}{2\pi}\frac{4\sin^2\left((u/2)\ln\left(t/t_c\right)\right)}{u^2|a(iu)|^2}+c,
\end{equation}
where $c$ is an undetermined constant containing the short-time contributions. Finally we note that, for large $t$, the integrand in Eq.~\eqref{eqsupp:almostmsd} is a fast-oscillating function, and thus it accumulates around its maximum at $u\simeq 0$. Approximating $a(iu)$ by its value at $u=0$ and integrating yield Eq.~\eqref{eq:MSDlargetimesattractive}. 

\begin{figure}[h!]
    \centering
    \includegraphics[width=0.6\linewidth]{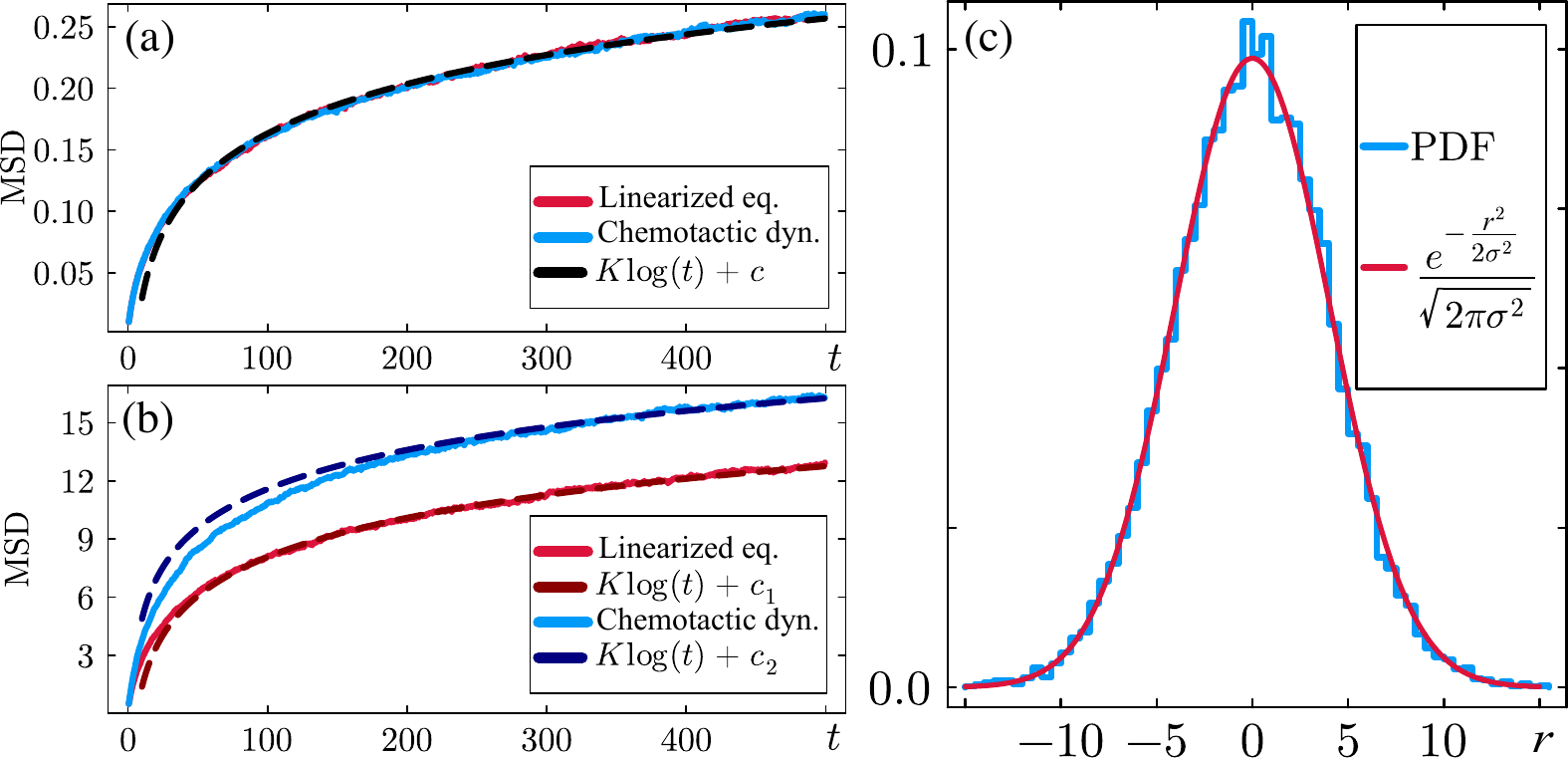}
    \caption{MSD of the particle with attractive chemotactic interaction at (a) low temperature $T=0.01$ or (b)  moderate values $T=0.5$ for $\alpha=-1.5$, $t_R=0$. The solid lines correspond to simulations of Eqs.~\eqref{eq:langevin_selfchemo} (blue) and~\eqref{eq:genlangevin} (red) while the dashed lines correspond to the analytical prediction according to Eq.~\eqref{eq:MSDlargetimesattractive}, with $K=(8\pi T)/(\alpha^2 a^2(0))$. The coefficients $c$, $c_1$ and $c_2$ are determined through a fit. (c) The probability distribution function (PDF) of $r(t)$ from Eq.~\eqref{eq:langevin_selfchemo} at large $t$ is a Gaussian with variance $\sigma^2(t)=\langle r^2(t)\rangle$ due to the linearity of equation~\eqref{eq:genlangevin}, confirming that the latter is a good approximation of~\eqref{eq:langevin_selfchemo}. 
    }
    \label{fig:1d_attractive}
\end{figure}

To confirm our findings, we simulate both Eqs.~\eqref{eq:langevin_selfchemo} and~\eqref{eq:genlangevin} for $\alpha<0$ at low and moderate temperatures, yielding the results shown in Fig.~\ref{fig:1d_attractive}.
As expected, the MSD for the two dynamics coincide at low $T$.  Upon increasing  $T$, the short-time behaviors differ, but both models converge to the same asymptotics in the limit  $t\rightarrow\infty$ of large times, which in all cases is well described by Eq.~\eqref{eq:MSDlargetimesattractive}. The probability distribution (PDF) of $r$ at large $t$ is Gaussian due to the linearity of Eq.~\eqref{eq:genlangevin}.

\section{Equation for $\delta r$ in reciprocal space}

In order to derive Eq.~\eqref{eq:lap_transf}, we decompose the chemotactic force as
\begin{equation}
    \label{eqsupp:chemodecomposition}
    -\alpha\nabla\phi(r_s(t),t)=\frac{\alpha}{4\sqrt{\pi}}\int_{t_{lt}}^t\frac{r_s(t)-r_s(t')}{(t-t')^{3/2}}\exp\left[-\frac{r_s(t)-r_s(t')}{4(t-t')}\right]+\fol(r_s(t),t),
\end{equation}
where $t_{lt}$ denotes the time of the last tumbling before $t$ and $\fol$ is due to the chemical field which has been produced before $t_{lt}$. We now substitute $r_s=\ra+\delta r$ in the first term on the r.h.s.~of Eq.~\eqref{eqsupp:chemodecomposition} and expand in $\delta r$:
\begin{align}
\label{eqsupp:stepinstability}
    &\frac{\alpha}{4\sqrt{\pi}}\int_{t_{lt}}^t\frac{r_s(t)-r_s(t')}{(t-t')^{3/2}}\exp\left[-\frac{r_s(t)-r_s(t')}{4(t-t')}\right]=\\
    &\frac{\alpha}{4\sqrt{\pi}}\int_{t_{lt}}^t\frac{\ra(t)-\ra(t')}{(t-t')^{3/2}}\exp\left[-\frac{\ra(t)-\ra(t')}{4(t-t')}\right]+\frac{\alpha}{4\sqrt{\pi}}\int_{t_{lt}}^t\frac{\delta r(t)-\delta r(t')}{(t-t')^{3/2}}\left[1-\frac{\ra(t)-\ra(t')}{2(t-t')}\right]\exp\left[-\frac{\ra(t)-\ra(t')}{4(t-t')}\right].\nonumber
\end{align}
Since $\dra$ is a slow function of $t$ we can evaluate Eq.~\eqref{eqsupp:stepinstability} by expanding $\ra(t')$ in Taylor series in $t'$ around $t$. In the first term on the r.h.s.~we expand up to the second order, i.e., $\ra(t')\simeq \ra(t)-\dra(t)(t-t')+\frac{1}{2}\ddra(t)(t-t')^2$, ignoring contributions of order higher than 2 in the time derivative. In the second term, instead, the expansion is considered only up to the first order, i.e., 
$\ra(t')\simeq \ra(t)-\dra(t)(t-t')$ since $\delta r$ is already vanishingly small in the adiabatic limit. 
Moreover, at sufficiently low $T$, $t_{lt}$ in the integrals can be replaced by $=-\infty$ because tumblings become really rare. After some algebraic manipulations, we find the equation for $\delta r$ to be:
\begin{equation}
    \label{eq:adiab_epansion}
    \dot{\delta r} (t)=\int_0^t dt' \gamma(t-t') \dot{\delta r}  (t') +\alpha     \frac{\ddra}{v_a^3}(t),
\end{equation}
with  $\gamma(t)=\frac{\alpha v_a}{4\sqrt{\pi}}\lt[\frac{1}{2}\Gamma(-1/2,v_a^2 t/4)-\Gamma(1/2, v_a^2 t/4)\rt]$, where $v_a=|\dra(t)|$, while $\Gamma(a,b)$ is the incomplete Gamma function. 
By taking the Laplace transform of both sides of Eq.~\eqref{eq:adiab_epansion} and by using the fact that $v_a$ can be treated as a constant while calculating the Laplace transform of $\gamma$, one obtains Eq.~\eqref{eq:lap_transf} of the main text.

\section{Derivation of the tumbling rate from the tumbling condition}

As discussed in the main text, the tumbling rate $\lambda^\epsilon(\fol)$ is the rate at which the stochastic force $\zeta_s(t)$ overcomes (in absolute value) $f_c+\epsilon  \fol$ while having a sign opposite to that of the self-propulsion speed of the particle. 
In order to compute $\lambda^\epsilon(\fol)$, we note that the slow noise $\zeta_s(t)$ --- defined in the main text as $\zeta_s(t) = \tau^{-1} \int_{-\infty}^t dt'\,e^{-(t-t')/\tau}\zeta(t')$, where $\tau$ is a coarse-graining time --- can be alternatively written as $\zeta_s(t)=\tau^{-1}\eta(t)$, where $\eta(t)$ is an Ornstein–Uhlenbeck (OU) process~\cite{uhlenbeck_theory_1930} with
\begin{equation}
    \dot \eta(t)=-\frac{1}{\tau}\eta(t)+\sqrt{2T}\zeta(t).
\end{equation}
The tumbling rate $\lambda^\epsilon(\fol)$ is thus the inverse of the mean-first passage time $\bar \tau$ of $\eta$ to surpass a barrier at $-\epsilon\tau(f_c+\epsilon \fol)$. 
At low $T$ this is given by~\cite{grebenkov_first_2014}
%

%
%
\begin{equation}
    \bar\tau=\frac{\sqrt{2\pi T \tau}}{(f_c+ \epsilon \fol)}\exp{\lt[ \frac{(f_c +\epsilon \fol)^2\tau}{2T}\rt]}.
\end{equation}
As explained in the main text, we choose $\tau$ to be proportional to the intrinsic timescale of the dynamics $\tau_\alpha$: $\tau = A \tau_\alpha \equiv 16 A /\alpha^2$, where $A$ is a fitting parameter. By using this definition and by introducing an additional fitting parameter $B$ for the overall amplitude, we find Eq.~\eqref{eq:tumblingrate}.

\section{Chemical field emitted by a particle moving with a constant velocity}
\label{sec:supp:profile-v}

We derive here the expression of the chemical field $\phi_+$ emitted by a particle moving at a constant velocity $v>0$. For convenience, we work in a reference frame comoving with the particle and centered such that the particle is at $x=0$. The chemical field satisfies the equation:
\begin{equation}
    \label{eqsupp:chemfield}
    -v\partial_x\phi_+(x)-\partial_x^2\phi_+(x)=\delta(x).
\end{equation}
For $x<0$ boundedness of $\phi_+$ requires $\partial_x\phi_+=0$, while for $x>0$ one has $\partial_x\phi_+=Ce^{-vx}$. To fix the value of $C$ we integrate Eq.~\eqref{eqsupp:chemfield} from $-\delta$ to $\delta$ for $\delta\rightarrow 0$. This give the condition: $\partial_x\phi_+(-\delta)-\partial_x\phi_+(\delta)=1$, which fixes $C=-1$ and gives:
\begin{equation}
    \label{eq:gradchem}
    \partial_x\phi_+(x)=
    \begin{cases}
        0&\quad \text{for} \quad x<0,\\
        -e^{-vx}&\quad \text{for} \quad x>0.
    \end{cases}
\end{equation}
The case of $v<0$ can be obtained analogously, finding $\partial_x\phi_-(x) = -\partial_x\phi_+(-x)$.
For $v$ which is small compared with the typical distance $l$ traveled by the particle, Eq.~\eqref{eq:gradchem} is strongly peaked around $x=0$ and, accordingly, $\partial_x\phi_+(x)$ can be approximated by a delta function, as reported in the main text.


\section{Mean-field equations and scaling for the run and tumble model}
\label{app:FPandscaling}

In this section we derive Eq.~\eqref{eq:FP_eff} for the run-and-tumble model. According to Eq.~\eqref{eq:old_field}, the maximum magnitude of the contribution of a tumbling event to $\fol$ scales as $\Delta t^{-1/2}$, where $\Delta t$ is the time elapsed since the tumbling event.
This means that the chemotactic force from the past trajectory is (at most) of order $\fol\simeq \sqrt{\lambda_0}$ and, at low $T$, remains small compared to the critical force $f_c \propto \tau_\alpha^{-1/2}$, allowing us to linearize the tumbling rates $\lambda^\pm(\fol)$ as $\lambda^\pm(\fol) = \lambda_0 \pm \lambda_1 \fol$. We then perform a mean-field approximation by replacing the field $\fol$ experienced by each particle with its ensemble average $\fol(r,t)\simeq \barf(r,t)$. At time $t$, $\barf$ is given by the second of Eq.~\eqref{eq:FP_eff}. 
Note that, while $\fol$ is the force due to the chemical field produced up to the last tumbling, we use the fact that, at sufficiently long time $t$, the time interval between two consecutive tumblings is negligible compared to $t$. This allows us to extend the time integral in the second of Eq.~\eqref{eq:FP_eff} to $t$. Consider now the probability $P_{\pm}(r,t)$ of having a particle at position $r$ and time $t$ traveling with velocity $\pm v$.  $P_{\pm}(r,t)$  evolves according to the master equation
\begin{equation}
\label{eqapp:ME}
    \partial_t P_{\pm}=\pm v\partial_rP_{\pm}+\lambda_0(P_{\mp}-P_{\pm})\pm\lambda_1(P_++P_-)\barf.
\end{equation}
We now rewrite this equation in terms of $P=P_++P_-$ and $\bars=P_+-P_-$, obtaining:
\begin{equation}
    \label{eqapp:before_enslaving}
    \begin{cases}
        \partial_tP=-v\partial_x\bars,\\
        \partial_t\bars=-v\partial_rP-2\lambda_0\bars-\lambda_1\barf P.
    \end{cases}
\end{equation}

As we are interested in scaling solutions, t large $t$ $\bars$ varies slowly as a function of $t$ and $\partial_t\bars\ll2\lambda_0s$.
Accordingly, the l.h.s.~of the second equation in Eq.~\eqref{eqapp:before_enslaving} vanishes and $\bars$ is enslaved to $P$ by the resulting equation $\bars=(-v\partial_r P-\lambda_1\barf P)/(2\lambda_0)$.

This equation, combined with the first of Eq.~\eqref{eqapp:before_enslaving} renders the first of Eq.~\eqref{eq:FP_eff}. In order to determine the scaling exponent of $\barf$ we then notice that this involves a convolution of $P$ with the Green’s function $G$, which is a function of $r/\sqrt{t}$: As we are searching for solutions of Eq.~\eqref{eq:FP_eff} with $z>1/2$, $P$ is significantly wider than $G$ for large $t$. This means that for $t\rightarrow\infty$, in the spatial integral which determines $\bar f$ according to Eq.~\eqref{eq:FP_eff}, 
$G$ can be approximated by a delta function. Using the scaling ansatz for $P(r,t)$ reported in the main text and involving the scaling function $U$, one can integrate in space the expression for $\bar f$ and, after changing the integration variable as $t' = u t$, one finds
\begin{equation}
\label{eq:fbar_scaling}
\bar{f}(r,t) = -\alpha t^{1 - 2z} \int_0^1 du\, u^z U'\left( \frac{r}{ut^z} \right).
\end{equation}
The remaining terms in Eq.~\eqref{eq:FP_eff} scale with time as $\partial_t P \sim t^{-z-1}$, $\partial_r^2 P \sim t^{-3z}$. 
Since $z > 1/2$, $\partial_r^2 P$ is negligible compared to $\partial_t P$ for  large $t$. Balancing the latter with the advective term yields $z=2/3$.

\section{Numerical methods}

 The simulations on the dynamics induced by Eq.~\eqref{eq:langevin_selfchemo} presented in the main text are obtained by numerical integration with an Euler scheme. At each timestep, the gradient of the chemical field at position $r(t)$ is computed from the integral representation given by the gradient of 
 Eq.~\eqref{eq:reg_gradient}. To discretize this integral, we split it into contributions from all previously recorded particle positions $r_i$ at discrete timesteps $i$.

Specifically, we approximate the integral as follows: we linearly interpolate the particle trajectory between two consecutive positions, $r_i$ and $r_{i+1}$, yielding a piecewise-linear representation and compute the contribution of each linear region separately. For numerical stability and accuracy, the integral contributions are computed differently for older and more recent time steps.

For older contributions (those occurring from timestep 0 up to timestep $i - m$), we employ a simple trapezoidal integration scheme.
For the last $m$ contributions, i.e., those closer in time to the current timestep, we utilize an adaptive Gaussian quadrature integration method.
In the simulations reported in the main text, we set $m = 100$. 

In the numerical simulations reported in Fig.~\ref{fig:attractive_diffusivities}(a) we set $dt=0.1$, and run $n=500$ instances of the discretized dynamics for $N=10^5$ steps for all but the last two points at $\alpha=0.913$ and $\alpha=1.513$. For these two, we reduce $dt$ by a factor $1/32$ to improve convergence close to the critical point. In Fig.~\ref{fig:attractive_diffusivities}(b) instead, we start from $dt=10^{-10}$ and run $n=1000$ instances of the dynamics. Then, we increase $dt$ by a factor of $1000$ and coarsen each trajectory by taking the position every $1000$ steps. Each of these trajectories is then used as an initial condition for the coarser dynamics at larger $dt$. We repeat this scaling procedure $6$ time to obtain the data for 
Fig.~\ref{fig:attractive_diffusivities}(b). For Fig.~\ref{fig:runandtumble}(a), we run our simulation at $dt=0.01$, $t_R=0$ for $N=10^5$ timesteps and $n=1000$ instances for each value of $\alpha$ and $T$. We acquire the first tumbling time $t_{\rm{tbl}}$,  defined as the first instance when the velocity, smoothed via a running average over a window $\tau$, changes sign. Trajectories that do not tumble during the simulation are assigned a tumbling time $t_{\rm{tbl}}$ equal to the simulation time $\bar t=N*dt$, i.e., $t_{\rm{tbl}}=\bar t$. Assuming that tumbling events follow a Poisson distribution with rate $\lambda_0$, the mean of $t_{\rm{tbl}}$ is given by
\begin{equation}
    \label{eqapp:ttbl}
    \langle t_{\rm{tbl}}\rangle=(1-e^{-\lambda_0 \bar t})/\lambda_0.
\end{equation}
We compute $\langle t_{\rm{tbl}}\rangle$ from simulation and use Eq.~\eqref{eqapp:ttbl} to estimate $\lambda_0$ in Fig.~\ref{fig:runandtumble}. The simulation of the dicretized chemoticatic dynamics for panels (b) and (c) of Fig.~\ref{fig:runandtumble} are performed with $dt=0.025$, $N=24000$, and $n=1000$. The effective run-and-tumble model is obtained as follows: at each timestep, the force $\fol(r(t),t)$ is calculated using Eq.~\eqref{eq:old_field}. A slow noise term $\zeta_s(t)$ is drawn at each step and the particle velocities are inverted whenever the noise exceeds the threshold $-\epsilon(\fol(r(t),t)+f_c)$, where $\epsilon=\pm 1$ is the sign of the current velocity. 

Finally, in Fig.~\ref{fig:1d_attractive}, both the original dynamics [see Eq.~\eqref{eq:langevin_selfchemo}] and its linearized approximation [ in Eq.~\eqref{eq:genlangevin}] are simulated using a timestep $dt=0.5$, with trajectories of $N=1000$ steps each and averaging over $n=15000$ instances.
 The linearized dynamics is obtained by approximating $\nabla G_1(r(t)-r(t'),t-t')\simeq -\frac{1}{4\sqrt{\pi}}\frac{r(t)-r(t')}{(t-t')^{3/2}}$ for the gradient in Eq.~\eqref{eq:reg_gradient}.

\end{document}